\documentclass{appolb}
\usepackage{epsfig}

\def\lQ{\Lambda_{\rm QCD}}
\newcommand{\nn}{\nonumber}
\newcommand{\be}{\begin{equation}}
\newcommand{\ee}{\end{equation}}
\newcommand{\bea}{\begin{eqnarray}}
\newcommand{\eea}{\end{eqnarray}}

\def\als{\alpha_{\rm s}}
\def\siml{{\
\lower-1.2pt\vbox{\hbox{\rlap{$<$}\lower6pt\vbox{\hbox{$\sim$}}}}\ }}  

\begin{document}
\title{Inclusive electromagnetic decays of the heavy quarkonium at 
next to leading log accuracy%
\thanks{Presented at XXVII International Conference of Theoretical
Physics, Ustron, 15-21 September 2003, Poland}%
}
\author{Antonio Pineda
\address{Dept. d'Estructura i Constituents de la
  Mat\`eria, U. Barcelona \\ Diagonal 647, E-08028 Barcelona,
  Catalonia, Spain}
}
\maketitle
\begin{abstract}
We show that perturbation theory may give reasonable numbers for the
decays of the bottomonium and charmonium ground states to $e^+e^-$ and
to $\gamma\gamma$. To reach this conclusion it is important to perform
the resummation of logs. In particular, we obtain the value
$\Gamma(\eta_b (1S) \rightarrow \gamma\gamma)=0.35 \pm 0.1 ({\rm th.})
\pm 0.05 (\lQ)$ KeV.
\end{abstract}
  
\section{Introduction}

In Ref. \cite{rgcsNLL}, the decays of Heavy Quarkonium to $e^+e^-$ and
$\gamma\gamma$ were computed with next to leading log (NLL) accuracy
within perturbation theory. When presenting these results in this
conference, one (reasonable) complaint was the complete absence of
numbers. We would like to fill this gap by doing the phenomenological
analysis of these results that we quote here for ease of reference.

We first quote the matching coefficients at the hard scale \cite{Barbieri,Barbieri2}:
\be
b_{\rm 1}(m)=1-2C_f{\als(m) \over \pi}
,
\ee
\be
b_0 (m)=1+\left({\pi^2 \over 4}-5\right){C_f \over 2}{\als(m) \over \pi}
,
\ee
where $C_f=(N_c^2-1)/(2N_c)$, whereas the renormalization group improved matching coefficients at
NLL (for the vector and pseudo-scalar) read (\cite{rgcsNLL,HS})
\bea 
&&B_s(\nu_p)=b_s(m)+ 
	A_1 {\als(m) \over w^{\beta_0}} \ln(w^{\beta_0})+ 
	A_2 \als(m) \bigg[z^{\beta_0}-1\bigg] 
\\
\nn
&&
+ 
	A_3 \als(m) \bigg[z^{\beta_0-2 C_A}-1  \bigg]+
	A_4 \als(m) \bigg[z^{\beta_0-13C_A/6}-1 \bigg]+
	A_5 \als(m) \ln(z^{\beta_0})
\,,
\label{cssln}
\eea
where $\beta_0={11 \over 3}C_A -{4 \over 3}T_Fn_f$, $z=\left[{\als(\nu_p) \over 
\als(m)}\right]^{1 \over
\beta_0}$ and $w=\left[{\als(\nu_p^2/m) \over \als(\nu_p)}\right]^{1 \over
\beta_0}$. The coefficients $A_i$ in Eq.~(\ref{cssln}) read
\begin{eqnarray} 
\label{Acoeffs}
  A_1 &=& {8\pi C_f \over 3\beta_0^2 }\left(C_A^2 +2C_f^2+3 C_fC_A
  \right) \,,\nn\\ 
  A_2 &=& { \pi C_f [3 \beta_0(26C_A^2+19C_A C_f-32C_f^2)-
     C_A(208 C_A^2+651 C_A C_f+116 C_f^2)]\over 78\,\beta_0^2\, C_A } \,, \nn\\
  A_3 &=& -{\pi C_f^2 \Big[  \beta_0 (4s(s+1)-3)+ C_A (15-14s(s+1)) \Big]
     \over 6 (\beta_0-2 C_A)^2 }\,,\nn\\
  A_4 &=& {24 \pi C_f^2 (3\beta_0-11 C_A)(5 C_A+8 C_f) \over 13\, C_A
     (6\beta_0-13 C_A)^2}\,,\nn\\
  A_5 &=& {-\pi C_f^2 \over \beta_0^2\, (6\beta_0-13C_A) (\beta_0-2C_A)}
      \, \Bigg\{ C_A^2(-9C_A+100C_f)
	\\ \nn 
	&&+\beta_0 C_A(-74C_f+C_A(42-13s(s+1)))
	+6\beta_0^2(2C_f+C_A(-3+s(s+1)))\Bigg\}
     \,.
\end{eqnarray}

By setting $\nu_p \sim mC_f\als/n$, $B_s(\nu_p)$ includes all the large logs at NLL order in 
any (inclusive enough) S-wave
heavy-quarkonium production observable we can think of. For instance, the decays to 
$e^+e^-$ and to two
photons at NLL order read
\bea
\label{vector}
\Gamma(V_Q (nS) \rightarrow e^+e^-) &=&
2{C_A \over 3} \left[ { \alpha_{em} Q \over M_{V_Q(nS)}} \right]^2
\left({m_QC_f\als \over n}\right)^3
\left\{
B_1(\nu_p)(1+\delta \phi_n)
\right\}^2
\\
\nn
&\simeq&
2 {C_A \over 3}
\left[ { \alpha_{em} Q \over M_{V_Q(nS)}} \right]^2
\left({m_QC_f\als \over n}\right)^3
\left\{1+2(B_1(\nu_p)-1)+2\delta \phi_n
\right\}
\,,
\\
\label{pseudo}
\Gamma(P_Q (nS) \rightarrow \gamma\gamma) &=&
2 C_A \left[ { \alpha_{em} Q^2 \over M_{P_Q(nS)}} \right]^2
\left({m_QC_f\als \over n}\right)^3
\left\{
B_0(\nu_p)(1+\delta \phi_n)
\right\}^2
\\
\nn
&\simeq&
2 C_A \left[ { \alpha_{em} Q^2 \over M_{P_Q(nS)}} \right]^2
\left({m_QC_f\als \over n}\right)^3
\left\{1+2(B_0(\nu_p)-1)+2\delta \phi_n
\right\}
\,,
\eea
where $V$ and $P$ stand for the vector and pseudo-scalar heavy
quarkonium, $\als=\als(\nu_p)$, and
($\Psi_n(z) ={d^n \ln \Gamma (z) \over dz^n}$ and $\Gamma (z)$ is the
Euler $\Gamma$-function)
\be
\delta \phi_n ={\als \over \pi}
\left[-C_A+
{\beta_0\over 4}
\left(3\log{\left({\nu_p n \over m C_f \als}\right)}+\Psi_1(n+1)-2n\Psi_2(n)+{3 \over 2}+\gamma_E+{2\over n}
\right)
\right]
\,,
\ee
which has been read from Ref. \cite{leadinglog}.

It is not our aim to perform a full fledge analysis of
Eqs. (\ref{vector}) and (\ref{pseudo}) here but rather to 
see what are the general trends obtained by the introduction of the
resummation of logs, as well as to give some predictions when the
results appear to be reliable enough.

\section{Phenomenological analysis}

In this section we perform the phenomenological analysis of the above
results for the bottomonium and charmonium systems.

\subsection{$b$-$\bar b$ 1S states}

We first consider the $b$-$\bar b$ 1S states and their decays
$\Gamma(\Upsilon (1S) \rightarrow e^+e^-)$ and $\Gamma(\eta_b (1S)
\rightarrow \gamma\gamma)$. For the first decay we will be able to
compare our results with experiment whereas our numbers for the second
can be considered to be a prediction.

\begin{figure}[h]
\hspace{-0.1in}
\epsfxsize=4.4in
\centerline{
\epsffile{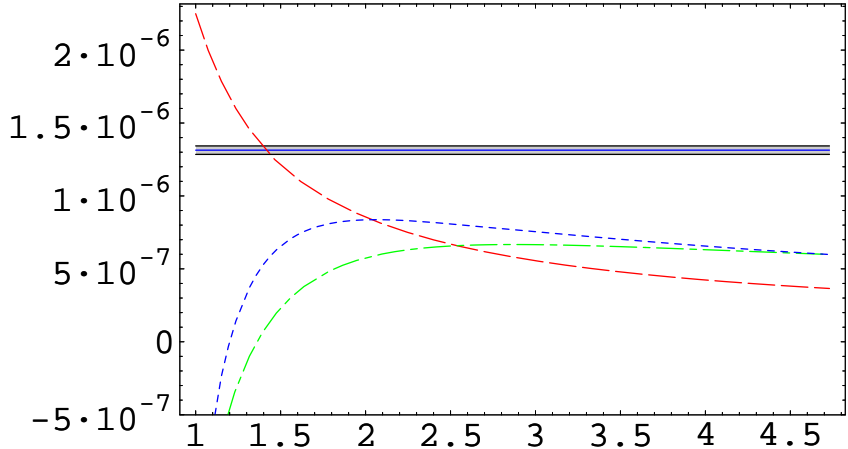}
}
\caption {{\it Plot of $\Gamma(\Upsilon (1S) \rightarrow e^+e^-)$ 
with LO/LL (dashed line), NLO (dot-dashed line) and NLL (dotted line) accuracy   
versus the renormalization scale $\nu$. The horizontal line and its
band give the experimental value and its errors: $\Gamma(\Upsilon (1S)
\rightarrow e^+e^-)=1.314 \pm 0.029$ KeV \cite{Hag}.}}
\label{Upsilon1See}
\end{figure}

We plot $\Gamma(\Upsilon (1S) \rightarrow e^+e^-)$ in
Fig. \ref{Upsilon1See} versus the renormalization scale $\nu$. We
consider the LO/LL result (they are equal), the NLO result and the NLL
result. We can see that the LO/LL result can match the experimental
figure for a reasonable value of $\nu$, close to the soft scale. This
agreement is destroyed once the NLO is considered. The best value we
can obtain is $0.667$ KeV. The reason seems to be the fact that we now
have two scales in the problem: the hard ($\sim m$) and the soft
($\sim m\als$) scale. The final outcome is that perturbation theory
breaks down before the normalization scale $\nu$ can reach the typical
soft scale of the problem. The standard solution to this problem goes
by doing a renormalization group analysis, summing up all the large
logs that appear in the problem. Actually, in our case, we are going
to have large logs produced by the ratio of the hard and soft scale
and by the ratio of the soft and ultrasoft scale ($\sim m\als^2$). We
can see that the use of the NLL expressions improves the agreement
with the experimental result (the best value now becomes $0.837$ KeV)
and enlarges the range of applicability of perturbation
theory. Moreover, the expansion seems to be convergent, being the NLL
result a correction with respect the LL order one.  Nevertheless,
perturbation theory still seems to break down before the
renormalization scale $\nu$ can reach the
typical soft scale (although getting closer to it than in the NLO
calculation) and the optimal result is off the experimental value by
around 50\%. This is a large effect. Therefore, in order to confirm
this picture, a full NNLL result should be obtained. This is a
difficult computation but were the convergent pattern confirmed the
outcomes will be important. One could then start quantifying the size
of the non-perturbative effects in a reliable manner.

We now perform a similar analysis for $\Gamma(\eta_b (1S) \rightarrow
\gamma\gamma)$.  For this decay no experimental figure
exists. Therefore, we will be able to give a prediction for it. The
picture is completely analogous to the $\Gamma(\Upsilon (1S)
\rightarrow e^+e^-)$ case. Actually the scale of minimal sensitive is
a little bit smaller than in the previous case, which makes us more
confident about the result. This confidence will be further boost
below when we consider the charmonium case for which experimental
numbers exist. We anticipate that very good agreement is reached in
that case. We then give a prediction for this decay:
\be
\label{etab}
\Gamma(\eta_b (1S) \rightarrow \gamma\gamma)=0.35 \pm 0.1 ({\rm th.})
\pm 0.05 (\lQ) {\rm KeV},
\ee
where ``$\lQ$'' stands for the error due to the variation of $\als$
($\alpha_s(M_Z)=0.118\pm 0.003$) and ``th'' for the theoretical
errors. The later are difficult to obtain and here we only pretend to
roughly estimate their size. We consider the variation of the decay if
we increase $\nu$ by two GeV with respect the optimal scale. This
gives the theoretical error quoted in Eq. (\ref{etab}). We consider
this estimate conservative in view of the good agreement with data
obtained for the $\eta_c$ case. Note that to have larger errors would
be inconsistent with the assumption that the remaining corrections
(perturbative and non-perturbative) are smaller, or at least of the
same order, than the difference between the LL and NLL result.
 
\begin{figure}[h]
\hspace{-0.1in}
\epsfxsize=4.4in
\centerline{
\epsffile{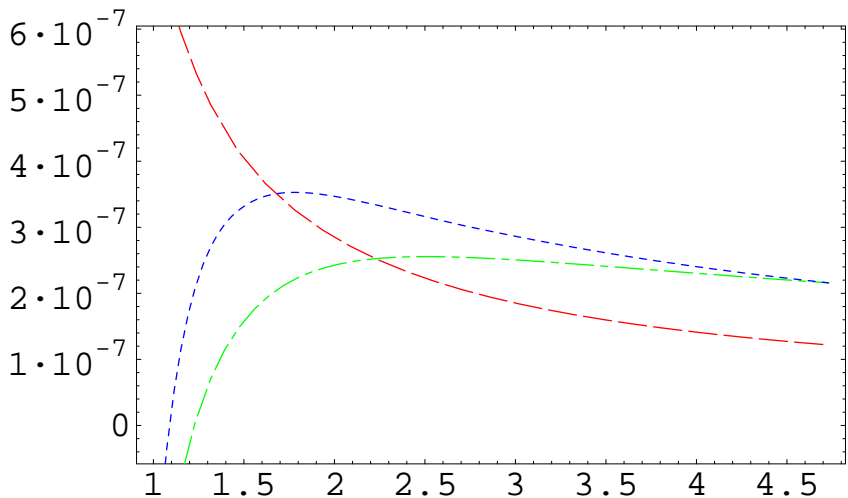}
}
\caption {{\it Plot of $\Gamma(\eta_b (1S) \rightarrow \gamma\gamma)$
with LO/LL (dashed line), NLO (dot-dashed line) and NLL (dotted line)
accuracy versus the renormalization scale $\nu$.}}
\label{etab1Sgg}
\end{figure}

For the above states one may consider reasonable to believe that a
perturbative approach is a good starting point for these systems,
since the soft scale is clearly in the perturbative
regime. Nevertheless, one should be careful since the ultrasoft scale
also enters the game through the matching coefficient (ultrasoft
effects first appear at $O(\als^3\ln^2\als)$). This means that for
$\nu \siml 1.5$ GeV, $\als(\nu^2/m_b)$ starts to blow up. Obviously,
the situation is even worse for the $n=2$ bottomonium states, since
the soft scale goes divided by $1/n^2$ (actually, we anticipate that a
bad behavior is found in this case). Surprisingly, however, for the
charmonium system we are in a situation quite similar to the $n=1$
bottomonium states since, although the typical soft scale is smaller
than the $n=1$ bottomonium soft scale, this is compensated by the fact
that the charm mass is smaller than the bottom one, so that we can run
down $\nu$ to scales quite close to the typical soft scales of the
problem before the break down of perturbation theory takes place. With
all these qualifications let us consider the $n=1$ charmonium or $n=2$
bottomonium states and see whether reasonable numbers are obtained.

\subsection{$n=1$ charmonium states}

\begin{figure}[h]
\hspace{-0.1in}
\epsfxsize=4.4in
\centerline{
\epsffile{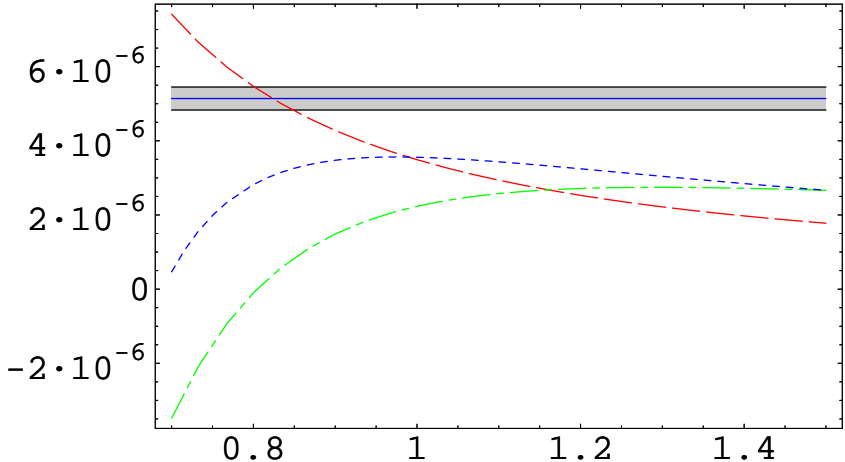}
}
\caption {{\it Plot of $\Gamma(J/\Psi (1S) \rightarrow e^+e^-)$ 
with LO/LL (dashed line), NLO (dot-dashed line) and NLL (dotted line) accuracy   
versus the renormalization scale $\nu$. The horizontal line and its
band give the experimental value and its errors: $\Gamma(J/\Psi (1S)
\rightarrow e^+e^-)=5.14 \pm 0.31$ KeV \cite{Hag}.}}
\label{Jpsi1See}
\end{figure}

We study $\Gamma(J/\Psi (1S) \rightarrow e^+e^-)$ in
Fig. \ref{Jpsi1See}.  Surprisingly a pretty similar picture than in
the $\Gamma(\Upsilon (1S) \rightarrow e^+e^-)$ appears. Actually, the
difference with experiment is of the same order, around 50 \%.

\begin{figure}[h]
\hspace{-0.1in}
\epsfxsize=4.4in
\centerline{
\epsffile{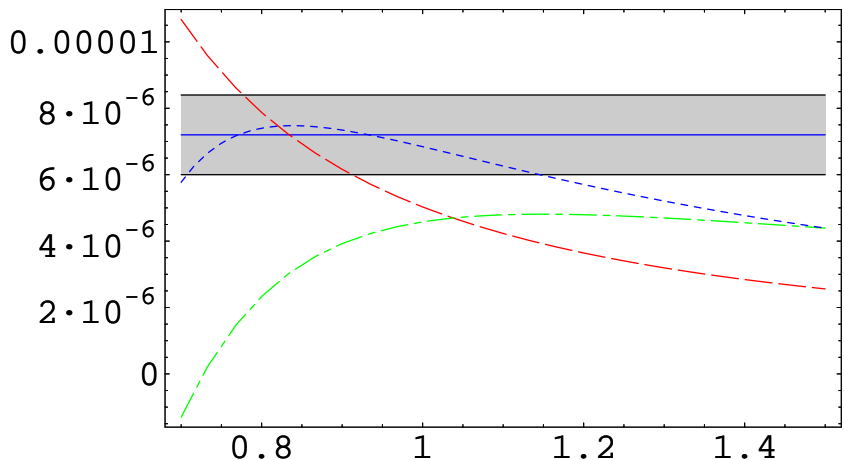}
}
\caption {{\it Plot of $\Gamma(\eta_c (1S) \rightarrow \gamma\gamma)$ 
with LO/LL (dashed line), NLO (dot-dashed line) and NLL (solid line) accuracy   
versus the renormalization scale $\nu$. The horizontal line and its
band give the experimental value and its errors: $\Gamma(\eta_c (1S) \rightarrow \gamma\gamma)=7.2 \pm 1.2$ KeV \cite{Hag}.}}
\label{etac1Sgg}
\end{figure}

We now consider $\Gamma(\eta_c (1S) \rightarrow \gamma\gamma)$ in Fig.
\ref{etac1Sgg}. In this case we have experimental data to compare
with, unlike the $\eta_b$ case. A similar pattern to the $\eta_b$
case is found and,
moreover, we get agreement with experiment. This is quite shocking
since, at the scale of minimal sensitive, the perturbative result hits
the experimental one just in the middle. Note that to get this
agreement the resummation of logs is necessary. The scale of minimal
sensitive is around $850$ MeV, of the same order than the typical soft
scale of the state. However, these numbers should be taken with caution
at this stage, since the running involves the ultrasoft scale, which
has been run down to very low scales $\sim$ 500 MeV in a correlated
way.

\subsection{$n=2$ bottomonium states}

\begin{figure}[h]
\hspace{-0.1in}
\epsfxsize=4.4in
\centerline{
\epsffile{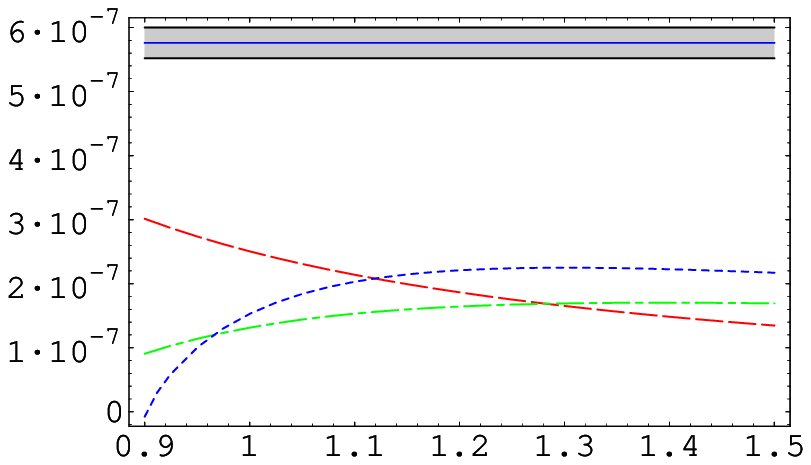}
}
\caption {{\it Plot of $\Gamma(\Upsilon (2S) \rightarrow e^+e^-)$ with
LO/LL (dashed line), NLO (dot-dashed line) and NLL (dotted line)
accuracy versus the renormalization scale $\nu$. The horizontal line and its
band give the experimental value and its errors: $\Gamma(\Upsilon (2S) \rightarrow e^+e^-)=0.576 \pm 0.024$ KeV \cite{Hag}.}}
\label{Upsilon2See}
\end{figure}

\begin{figure}[h]
\hspace{-0.1in}
\epsfxsize=4.4in
\centerline{
\epsffile{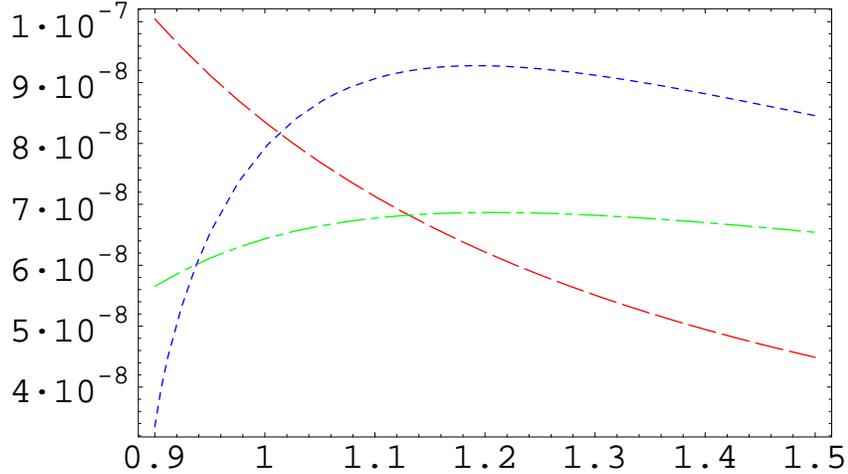}
}
\caption {{\it Plot of $\Gamma(\eta_b (2S) \rightarrow \gamma\gamma)$
with LO/LL (dashed line), NLO (dot-dashed line) and NLL (dotted line)
accuracy versus the renormalization scale $\nu$.}}
\label{etab2Sgg}
\end{figure}

We now turn to the $n=2$ bottomonium states. We show our results in
Figs. \ref{Upsilon2See} and \ref{etab2Sgg}. In principle, this is the
most problematic case since the soft scale of the $\Upsilon(2S)$ is
$\sim 1/n^2(=1/4)$ $\times$ the soft scale of the $\Upsilon(1S)$ (even if
partially corrected by the fact that $\als$ would be larger for the 2S
state than for the 1S state). Actually for $\Gamma(\Upsilon (2S) \rightarrow e^+e^-)$ we can
compare with experiment and our result is a factor two smaller than
the experimental number. We note that, even for the leading order result, we
have to go to very small scales to get agreement with experiment
($\sim 600$ MeV). This raises doubts about our perturbative
analysis for the $n=2$ bottomonium states. Somewhat, even if the
resummation of logs helps, perturbation theory still breaks down
before one can reach the typical soft scale of the problem (which is
very small). 

\subsection{Final discussion}

Even if one should wait for the complete NNLL result, we can start to
see some trends. The uncalculated (perturbative or non-perturbative)
terms seems to be larger for the decays into $e^+e^-$ than for the
decays into two photons. One could then start to speculate about the
size of the non-perturbative effects in each case. Good enough, the
agreement obtained with the experimental figure of $\Gamma(\eta_c (1S)
\rightarrow \gamma\gamma)$ makes us quite confident that the result we
have obtained for $\Gamma(\eta_b (1S) \rightarrow \gamma\gamma)$ will
be quite close to the experimental number.

\section{Conclusions}

We have performed a phenomenological analysis of the NLL results
obtained in Ref. \cite{rgcsNLL} for the heavy quarkonium decays to
$e^+e^-$ and to two photons. 

For $\Gamma(\Upsilon(1S) \rightarrow e^+e^-)$, we find that the
resummation of logs significantly improves the agreement of the
perturbative result with experiment in relation with a pure NLO
evaluation, such that, at the place of minimal sensitive, the
theoretical number is off the experimental one by around
50\%. Moreover, overall convergence is found, with the NLL order
result being a correction with respect the LL order one. Note that in
order to be so, we have to take $\nu$ of the order of the soft scale
such that the large logs are resummed. Surprisingly, the very same
picture holds for $\Gamma(J/\Psi (1S) \rightarrow e^+e^-)$ too. It
would certainly be challenging to have the complete NNLL result to
check whether this pattern of convergence survives, since the
difference with experiment is still large in both cases. If this is so
one could start to reliable estimate non-perturbative effects in heavy
quarkonium (and its relation with the ultrasoft cutoff) and, for the
charmonium case, support the view that one can actually use
perturbation theory as an starting point for its study (see
\cite{BSV}). On the other hand, for $\Gamma(\Upsilon(2S) \rightarrow
e^+e^-)$, we find an strong discrepancy with the experimental figure,
rasing doubts that one can actually use perturbation theory
there. This could make sense since, for Coulomb-type bound states, the
soft scale of the $\Upsilon(2S)$ would be $\sim 1/n^2(=1/4)$ $\times$
the soft scale of the $\Upsilon(1S)$ (partially corrected by the fact
that $\als$ would be larger for the 2S state than for the 1S one). Even if
the resummation of logs helps, perturbation theory still breaks down
before the renormalization scale $\nu$ can reach the typical soft scale of the problem. Somewhat
we feel that something similar still happens for the $\Upsilon(1S)$
and the $J/\Psi(1S)$ in a less severe way, since in these last cases
one can get much closer to what we believe are the typical soft scales
of the problem. In any case, we would not like to draw definite
conclusions just from this analysis.

For the decay of the heavy quarkonium to two photons we only have
experimental data for the Charmonium case. In this case perfect
agreement with experiment is obtained. This boosts our confidence that 
our number for $\Gamma(\eta_b (1S) \rightarrow \gamma\gamma)$ will be
reliable, which we take as one of the major results of this paper.

For $\Gamma(\eta_b (2S) \rightarrow \gamma\gamma)$, we prefer not to
draw any conclusion in view of the failure of the results obtained for
$\Gamma(\Upsilon(2S) \rightarrow e^+e^-)$.

It is a general trend that the resummation of logs improves the
agreement with the experimental result (when available). This can be
understood by the fact that makes perturbation theory stable up to
smaller scales. 

We can also see that the picture is quite similar to the one obtained
in Ref. \cite{KPPSS} for the hyperfine of the heavy quarkonium, which has
recently been computed with NLL accuracy.
 
The fact that we obtain reasonable numbers for the charmonium system
can be considered a surprise. One may think that in this case one has
run down to very low scales the ultrasoft scale. Actually, at the
numerical level, the ultrasoft scales for the ground states of the
bottomonium and charmonium seem to be similar. The reason is that,
even if the soft scale of charmonium is smaller than of bottomonium, this is
compensated by the fact that the charm mass is smaller than the bottom
mass. This
explains the similar behavior found for both systems with errors of
the same size. For the $n=2$ bottomonium states, the
behavior seems to be different, with a major breakdown of perturbation
theory, making the results not trustworthy (actually a factor two
disagreement with experiment is obtained when experimental results are
available).  In any case, even for the bottomonium and charmonium
ground state, the ultrasoft scale has been run down to very low
scales. This is a potential problem of the whole analysis. The issue
would be certainly clarified if a complete NNLL computation were
available for the decays. In case a convergent pattern is observed for
the perturbative series, it would certainly be considered an
indication that perturbation theory can be applied for these
systems. One can then start considering a quantitative study of
non-perturbative effects and its relation with the ultrasoft
cutoff. In particular one may start to consider what the ratio between
$\lQ$ and $mv^2$ exactly is and try to apply the results obtained in
Ref. \cite{NP} for the non-perturbative corrections.

\medskip

\noindent {\bf Acknowledgments}\\ I would like to thank the organizers
for the warm hospitality and the perfect organization of this
conference. The author is supported by MCyT and Feder (Spain),
FPA2001-3598, by CIRIT (Catalonia), 2001SGR-00065 and by the EU
network EURIDICE, HPRN-CT2002-00311.


\begin{thebibliography}{99}

\bibitem{rgcsNLL} A. Pineda,
Phys.\ Rev.\ D {\bf66}, 054022 (2002).

\bibitem{Barbieri} R. Barbieri, R. Gatto, R. Kogerler and Z. Kunszt,
Phys. Lett. {\bf B57}, 455 (1975). 

\bibitem{Barbieri2} I. Harris and L.M. Brown, Phys. Rev. {\bf 105}, 1656
(1957); R. Barbieri, E. d'Emilio, G. Curci and E. Remiddi,
Nucl. Phys. {\bf B154}, 535 (1979).

\bibitem{HS} A.H. Hoang and I.W. Stewart, 
Phys.\ Rev.\ D {\bf 67}, 014020 (2003).
 
\bibitem{leadinglog} K. Melnikov and A. Yelkhovsky, Phys. Rev. {\bf D59},
114009 (1999); A.A. Penin and A.A. Pivovarov, Nucl. Phys. {\bf B549},
217 (1999).   

\bibitem{Hag} K. Hagiwara {\it et al.},
Phys.\ Rev.\ D {\bf 66}, 010001 (2002).

\bibitem{BSV} N. Brambilla, Y. Sumino and A. Vairo, Phys. Lett. {\bf
B513}, 381 (2001); S. Recksiegel and Y. Sumino, hep-ph/0305178. 

\bibitem{KPPSS} B.A. Kniehl, A.A. Penin, A. Pineda, V.A. Smirnov and
M. Steinhauser, hep-ph/0312086.

\bibitem{NP} N. Brambilla, D. Eiras, A. Pineda, A. Vairo and J. Soto, 
Phys. Rev. {\bf D67}, 034018 (2003); 
N. Brambilla, A. Pineda, A. Vairo and J. Soto, hep-ph/0307159. 


\end{thebibliography}
\end{document}